\documentclass[twocolumn,prl,letterpaper,showpacs,floatfix]{revtex4}

\usepackage{graphicx}
\usepackage{dcolumn}
\usepackage{bm}
\usepackage{amsmath}

\begin{document}

\title{
Finite electron crystallites in strong magnetic fields: 
Precursors of a supersolid?
}

\author{Yuesong Li}
\author{Constantine Yannouleas}
\author{Uzi Landman}

\affiliation{School of Physics, Georgia Institute of Technology,
             Atlanta, Georgia 30332-0430}

\date{27 February 2005}

\begin{abstract}
We show that a supersolid phase, exhibiting simultaneously solid and superfluid 
behavior, properly describes the finite electron crystallites that form in 
two-dimensional quantum dots under high magnetic fields. These crystallites 
rotate already in their ground state and exhibit a nonclassical rotational 
inertia. They are precursors to a supersolid crystal in the lowest Landau level.
We use exact numerical diagonalization, calculations employing analytic
many-body wave functions, and a newly derived analytic expression for the
total energies that permits calculations for arbitrary number of electrons.
\end{abstract}

\maketitle

The existence of an exotic supersolid crystalline phase with combined solid and 
superfluid characteristics has been long conjectured \cite{anlf,ches,legg}. 
This conjecture was made in relation to solid $^4$He. The recent experimental 
discovery \cite{chan} that solid $^4$He exhibits a nonclassical (nonrigid) 
rotational inertia (NCRI \cite{legg}) has revived an intense interest 
\cite{legg2,datt,prok,cepe,fala} in the existence and properties of the 
supersolid phase.

The supersolid phase is not restricted to bosonic crystals \cite{fala,kato}. In 
this paper, we demonstrate the existence of such a phase in two-dimensional (2D) 
semiconductor $N$-electron quantum dots (QDs) under strong perpendicular 
magnetic fields ($B$). Central to the concept of a supersolid is the appearance 
of a NCRI when the crystal is made to rotate \cite{legg,legg2}. 

Under a high magnetic field, the electrons confined in a QD localize at the
vertices of concentric polygonal rings and form a rotating electron crystallite 
(REC) \cite{yl1,note1}. The ground-state many-electron wave function has a 
nonzero good total angular momentum quantum number. We show that the 
corresponding rotational inertia strongly deviates from the rigid classical 
value, a fact that endows the REC with supersolid characteristics. The REC at 
high $B$ can be viewed as the precursor of a supersolid crystal that develops in
the lowest Landau level (LLL) in the thermodynamic limit. These conclusions were
enabled by the development of an analytic expression for the energy of the REC 
that permits calculations for an arbitrary number of electrons, given the 
classical polygonal-ring structure in the QD \cite{kon}.

For strong $B$ we can approximate \cite{yl2} the single-electron wave function 
by (parameter free) displaced Gaussian functions; namely, for an electron 
localized at ${\bf R}_j$ ($Z_j$), we use \cite{yl2} the expression
\begin{equation}
u(z,Z_j) = \frac{1}{\sqrt{\pi} \lambda}
\exp \left( -\frac{|z-Z_j|^2}{2\lambda^2} - i\varphi(z,Z_j;B) \right),
\label{uhfo}
\end{equation}
with $\lambda = \sqrt{\hbar /m^* \Omega}$;
$\Omega=\sqrt{\omega_0^2+\omega_c^2/4}$, where $\omega_c=eB/(m^*c)$ is the
cyclotron frequency and $\omega_0$ specifies the external parabolic confinement
defining the QD. The position variables are given by the complex number 
$z=x+iy$ and $Z_j = X_j +i Y_j$.
The phase guarantees gauge invariance in the presence of
a perpendicular magnetic field and is given in the symmetric gauge by
$\varphi(z,Z_j;B) = (x Y_j - y X_j)/2 l_B^2$, with $l_B = \sqrt{\hbar c/ e B}$
being the magnetic length.

For an extended 2D system, the $Z_j$'s form a triangular lattice 
\cite{yosh}. For finite $N$, however, the $Z_j$'s coincide \cite{yl1,yl2,yl3} 
with the equilibrium positions [forming $r$ concentric 
regular polygons denoted as ($n_1, n_2,...,n_r$)] of $N=\sum_{q=1}^r n_q$ 
classical point charges inside an external parabolic confinement \cite{kon}. 
The wave function of the {\it static\/} electron crystallite (SEC) is a 
{\it single\/} Slater determinant $|\Psi^{\text{SEC}} [z] \rangle$ made out of 
the single-electron wave functions $u(z_i,Z_i)$, $i = 1,...,N$. 

$|\Psi^{\text{SEC}} [z] \rangle$ represents a broken-symmetry state, since it is
not an eigenstate of the total angular momentum operator with eigenvalue $L$. 
For a {\it rotating\/} electron crystallite, $L$ 
must be a good quantum number. To decsribe the rotation of the electron 
crystallite, we project the Slater determinant onto a state with good $L$ 
\cite{yl1,yl2,yl3,note2}, thus restoring the circular symmetry and obtaining
\begin{eqnarray}
|\Phi^{\text{REC}}_L \rangle && =  \int_0^{2\pi} ... \int_0^{2\pi}
d\gamma_1 ... d\gamma_r \nonumber \\
&& \times |\Psi^{\text{SEC}}(\gamma_1, ..., \gamma_r) \rangle
\exp \left( i \sum_{q=1}^r \gamma_q L_q \right).
\label{wfprj}
\end{eqnarray}
Here $L=\sum_{q=1}^r L_q$ and
$|\Psi^{\text{SEC}}[\gamma] \rangle$ is the original Slater determinant 
with {\it all the single-electron wave functions of the $q$th ring\/} rotated 
(collectively, i.e., coherently) by the {\it same\/} azimuthal angle $\gamma_q$.
Note that Eq.\ (\ref{wfprj}) can be written as a product of projection 
operators acting on the original Slater determinant [i.e., on 
$|\Psi^{\text{SEC}}(\gamma_1=0, ..., \gamma_r=0) \rangle$]. 
Setting $\lambda = l_B \sqrt{2}$ restricts the single-electron wave function in 
Eq. (\ref{uhfo}) to be entirely in the lowest Landau level \cite{yl1}.
The continuous-configuration-interaction form of the projected wave functions
[i.e., the linear superposition of determimants in Eq.\ (\ref{wfprj})]
implies a highly entangled state. We require here that $B$ is sufficiently
strong so that all the electrons are spin-polarized and that the ground-state 
angular momentum $L \geq L_0 \equiv N(N-1)/2$ [the minimum value $L_0$ specifies
the socalled maximum density droplet].

The energy of the REC state [Eq.\ (\ref{wfprj})] is given by \cite{yl2,yl3}
\begin{equation}
E^{\text{REC}}_L = \left. { \int_0^{2\pi} h([\gamma]) e^{i [\gamma] \cdot [L]}
d[\gamma] } \right/%
{ \int_0^{2\pi} n([\gamma]) e^{i [\gamma] \cdot [L]} d[\gamma]},
\label{eproj}
\end{equation}
with $h([\gamma]) =
\langle \Psi^{\text{SEC}}([0]) | H | \Psi^{\text{SEC}}([\gamma]) \rangle$,
$n([\gamma]) =
\langle \Psi^{\text{SEC}}([0]) | \Psi^{\text{SEC}}([\gamma]) \rangle$, and
$[\gamma] \cdot [L] = \sum_{q=1}^r \gamma_q L_q$.
The SEC energies are simply given by $E_{\text{SEC}} = h([0])/n([0])$.
The many-body Hamiltonian is 
\begin{equation}
H=\sum_{i=1}^N \frac{1}{2m^*}\left( {\bf p}_i- \frac{e}{c} {\bf A}_i \right)^2
+ \sum_{i=1}^N \frac{m^*}{2} \omega_0^2 {\bf r}_i^2
+ \sum_{i=1}^N \sum_{j > i}^N \frac{e^2}{\kappa r_{ij}},
\label{mbh}
\end{equation}
which describes $N$ electrons (interacting via a Coulomb repulsion) confined by
a parabolic potential of frequency $\omega_0$ and subjected to
a perpendicular magnetic
field $B$, whose vector potential is given in the symmetric gauge by
${\bf A(r)} = \frac{1}{2} (-By,Bx,0)$.
$m^*$ is the effective electron mass, $\kappa$ is the  dielectric
constant of the semiconductor material, and $r_{ij}=|{\bf r}_i - {\bf r}_j|$.
For sufficiently high magnetic fields, the electrons are fully 
spin-polarized and the Zeeman term (not shown here) does not need to be 
considered.

\begin{figure}[t]
\centering\includegraphics[width=6.3cm]{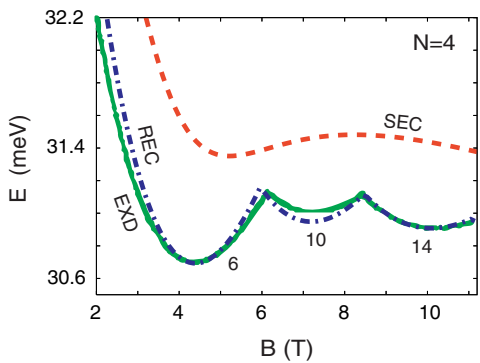}
\caption{
Ground-state energies calculated with the REC wave function compared to EXD 
calculations. Energies for $N=4$ fully polarized electrons (referenced to 
$4 \hbar \Omega$) as a function of the magnetic field $B$.
Dashed line (online red): Broken-symmetry SEC.
Solid line (online green): EXD (from Ref.\ \cite{ruan2}).
Dashed-dotted line (online blue): Symmetry-restored REC.
Numbers under the curve denote the value of magic angular momenta $L_m$ of the
ground state. Corresponding fractional filling factors are specified by
$\nu = N(N-1)/(2L_m)$.
Parameters used: confinement $\hbar \omega_0=3.60$ meV, 
dielectric constant $\kappa=13.1$, effective mass $m^*=0.067 m_e$. 
}
\end{figure}


Prior to demonstrating that the REC wave function [Eq.\ (\ref{wfprj})]
exhibits supersolid characteristics, we address the numerical accuracy 
of its energies [$E^{\text{REC}}_L$ in Eq.\ (\ref{eproj})] 
compared to exact diagonalization (EXD) solutions of the many-body problem
defined by the Hamiltonian in Eq.\ (\ref{mbh}). In Fig.\ 1, 
the ground-state energies as a function of $B$ are compared to EXD calculations
\cite{ruan2} for a QD with $N=4$ electrons. The thick dotted line (online red) 
represents the broken-symmetry SEC approximation, varying smoothly with $B$
and lying above the EXD curve [solid (online green)]. The REC curve 
[dashed-dotted (online blue)] agrees well with the EXD one in the entire range 
2 T$ < B <$ 11 T. The REC and EXD ground-state energies exhibit oscillations as 
a function of $B$. These oscillations reflect the incompressibility of the 
many-body states associated with magic angular momenta $L_m$. The $L_m$'s
are specified by the number of electrons on each polygonal ring; in 
general $L_m = L_0 + \sum_{q=1}^r k_q n_q$, with $k_q$ being
nonnegative integers, i.e., $L_m=6+4k_2$ for the (0,4) structure of $N=4$,
see Fig.\ 1. 

As a second accuracy test, we compare in TABLE I REC and EXD results for the 
interaction energies of the yrast band for $N=6$ electrons in the lowest Landau 
level (an yrast state is the lowest energy state for a given $L_m$). The
relative error is smaller than \%0.3, and it decreases steadily for larger $L$ 
values. The total energy of the REC is lower than that of 
the SEC, Fig.\ 1. A theorem discussed in Sec. III of Ref. \cite{low}, pertaining
to the energies of projected wave functions, guarantees that this lowering of
energy applies for all values of $N$ and $B$.

\begin{table}[b]
\caption{%
Comparison of yrast-band energies obtained from REC and EXD calculations for 
$N=6$ electrons in the lowest Landau level, that is in the
limit $B \rightarrow \infty$. In this limit the external confinement can
be neglected and only the interaction energy contributes to the yrast-band
energies. Energies in units of $e^2/(\kappa l_B)$. For the REC results, the
(1,5) polygonal-ring arrangement was considered. For $L < 140$, see
TABLE IV in Ref.\ [11(b)] and Table III in Ref.\ \cite{yl4}.
}
\begin{ruledtabular}
\begin{tabular}{rccc}
$L$  & REC & EXD  & Error (\%)   \\ \hline
140  & 1.6059  & 1.6006  & 0.33 \\
145  & 1.5773  & 1.5724  & 0.31 \\
150  & 1.5502  & 1.5455  & 0.30 \\
155  & 1.5244  & 1.5200  & 0.29 \\
160  & 1.4999  & 1.4957  & 0.28 \\
165  & 1.4765  & 1.4726  & 0.27 \\
170  & 1.4542  & 1.4505  & 0.26 \\
175  & 1.4329  & 1.4293  & 0.25 \\
180  & 1.4125  & 1.4091  & 0.24 \\
185  & 1.3929  & 1.3897  & 0.23 \\
190  & 1.3741  & 1.3710  & 0.23 \\
195  & 1.3561  & 1.3531  & 0.22 \\
200  & 1.3388  & 1.3359  & 0.21
\end{tabular}
\end{ruledtabular}
\end{table}

The crystalline polygonal-ring arrangement $(n_1,n_2,...,n_r)$ of classical 
point charges is portrayed directly in 
the electron density of the broken-symmetry SEC, since the latter consists of 
humps centered at the localization sites $Z_j$'s ({\it one hump} for each 
electron). In contrast, the REC has good angular momentum and thus its electron 
density is circularly uniform. To probe the crystalline character of the REC, 
we use the conditional probability distribution (CPD) defined as
\begin{equation}
P({\bf r},{\bf r}_0) =
\langle \Phi | 
\sum_{i \neq j}  \delta({\bf r}_i -{\bf r})
\delta({\bf r}_j-{\bf r}_0) 
| \Phi \rangle  / \langle \Phi | \Phi \rangle,
\label{cpds}
\end{equation} 
where $\Phi ({\bf r}_1, {\bf r}_2, ..., {\bf r}_N)$ 
denotes the many-body wave function under consideration.
$P({\bf r},{\bf r}_0)$ is proportional to the conditional probability of
finding an electron at ${\bf r}$, given that another electron 
is assumed at ${\bf r}_0$. This procedure subtracts the collective rotation 
in the laboratory frame of referenece, and, as a result, the 
CPDs reveal the structure of the many body state in the intrinsic (rotating) 
reference frame.

\begin{figure}[t]
\centering\includegraphics[width=8.cm]{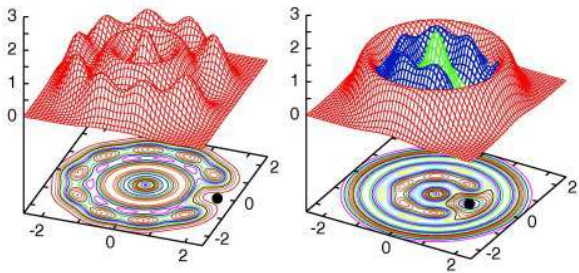}
\caption{
CPDs for the ground state of $N=17$ electrons at $B=10$ T ($L=228$). 
The observation point (solid dot) is placed (i) on the outer ring at 
$r_0=1.858R_0$, left frame, and (ii) on the inner ring at $r_0=0.969R_0$,
right frame. The rest of the parameters are the same as in Fig.\ 1.
Lengths in units of $R_0 =( 2e^2/(\kappa m^* \omega_0^2) )^{1/3}$. 
CPDs (vertical axes) in arbitrary units.
}
\end{figure}

In Fig.\ 2, we display the CPD for the REC wave function of $N=17$ electrons.
This case has a nontrivial three-ring structure (1,6,10) \cite{kon}, 
which is sufficiently complex to allow generalizations for larger numbers of 
particles. The remarkable combined character (partly crystalline 
and partly fluid leading to a NCRI) of the
REC is illustrated in the CPDs of Fig.\ 2. Indeed, as the two CPDs 
(reflecting the choice of taking the observation point [${\bf r}_0$ in 
Eq.\ (\ref{cpds})] on the outer, left frame, or the inner ring, right frame) 
reveal, the polygonal electron rings rotate {\it independently\/} of each other.
Thus, e.g., to an observer located on the inner ring, the outer ring will appear
as uniform, and vice versa. The wave functions obtained from exact 
diagonalization exhibit also the property of independently rotating rings 
(see e.g., the $N=9$ case in Ref.\ \cite{yl3}), which 
is a testimony to the ability of the REC wave function to capture the essential 
physics of a finite number of electrons in high $B$.

In addition to the conditional probabilities, the solid/fluid character of the 
REC is revealed in its excited rotational spectrum for a given $B$. From our 
microscopic calculations based on the wave function in Eq.\ (\ref{wfprj}), we 
have derived (see below) an approximate (denoted as ``app''), but 
{\it analytic\/} and {\it parameter-free\/}, expression [see Eq.\ 
(\ref{app})] which reflects directly the nonrigid (nonclassical) 
character of the REC for arbitrary size. This expression allows calculation of
the energies of RECs for arbitrary $N$, given the corresponding equilibrium 
configuration of confined classical point charges. 

We focus on the description of the yrast band at a given $B$.
Motivated by the aforementioned nonrigid character of the rotating electron
crystallite, we consider the following kinetic-energy term appropriate for a 
$(n_1,...,n_q,...,n_r)$ configuration (with $\sum_{q=1}^r n_q =N$):
\begin{equation}
E_{\text{app}}^{\text{kin}} (N)=
\sum_{q=1}^r \hbar^2 L_q^2/(2 {\cal J}_q (a_q)) - \hbar \omega_c L/2,
\label{eclkin}
\end{equation}
where $L_q$ is the partial angular momentum associated with the $q$th ring and
the total $L=\sum_{q=1}^r L_q$. ${\cal J}_q (a_q)) \equiv n_q m^* a_q^2$ is the 
rotational inertia of each {\it individual\/} ring, i.e., the moment of inertia
of $n_q$ classical point charges on the $q$th polygonal ring of radius $a_q$. 
To obtain the total energy, $E_L^{\text{REC}}$, we include also the term 
$E_{\text{app}}^{\text{hc}} (N) =\sum_{q=1}^r  {\cal J}_q (a_q) \Omega^2/2$
due to the effective harmonic confinement $\Omega$ [see discussion of Eq.\
(\ref{uhfo})], as well as the interaction energy $E_{\text{app}}^C$,
\begin{equation}
E_{\text{app}}^C (N) = \sum_{q=1}^r \frac{n_q S_q}{4} \frac{e^2}{\kappa a_q} +
\sum_{q=1}^{r-1} \sum_{s > q}^r V_C(a_q,a_s).
\label{vc}
\end{equation}
The first term is the Coulomb repulsion of $n_q$ point-like electrons
on a given ring, with $S_q = \sum_{j=2}^{n_q}(\sin[(j-1)\pi/n_q])^{-1}$,
and the second term is the Coulomb repulsion between (uniform) rings with
$V_C(a_q,a_s)= n_q {n_s} {_2F_1} [3/4,1/4;1;4 a_q^2 a_s^2(a_q^2+a_s^2)^{-2}]
e^2 (a_q^2+a_s^2)^{-1/2}/\kappa$, where ${_2F_1}$ is the
hypergeometric function.

For large $L$ (and/or $B$), the radii of the rings of the rotating crystallite 
can be found by neglecting the interaction term in the total approximate energy, 
thus minimizing only 
$E_{\text{app}}^{\text{kin}} (N) + E_{\text{app}}^{\text{hc}} (N)$.
One finds $a_q = \lambda \sqrt{L_q/n_q}$; the ring radii depend on $L_q$,
reflecting the {\it lack of radial rigidity\/}. Substitution into the above
expressions for $E_{\text{app}}^{\text{kin}}$, $E_{\text{app}}^{\text{hc}}$, and
$E_{\text{app}}^C$ yields for the total approximate energy the final
expression:
\begin{eqnarray}
E_{\text{app},L}^{\text{REC}}&&(N) =  \hbar(\Omega-\omega_c/2) L + \nonumber \\
&& \sum_{q=1}^r \frac{C_{V,q}}{L_q^{1/2}} + 
\sum_{q=1}^{r-1} \sum_{s > q}^r 
V_C(\lambda \sqrt{\frac{L_q}{n_q}}, \lambda \sqrt{\frac{L_s}{n_s}}), 
\label{app}
\end{eqnarray}
where the constants $C_{V,q}=0.25 n_q^{3/2} S_q e^2/(\kappa \lambda)$.
In the case of the simplest $(0,N)$ and $(1,N-1)$ ring configurations,
Eq.\ (\ref{app}) reduces to the partial expressions reported earlier 
\cite{yl2,mak}.

\begin{figure}[t]
\centering\includegraphics[width=7.5cm]{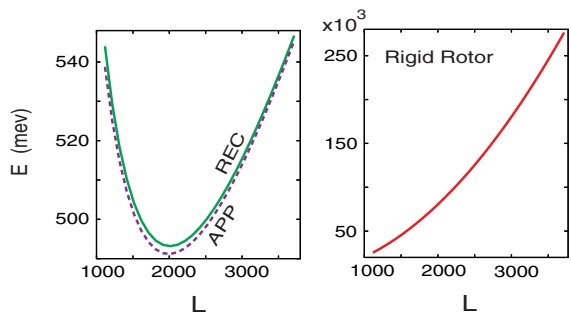}
\caption{
Left: Approximate analytic expression [Eq.\ (\ref{app}), dashed line (online 
violet)] compared with microscopic REC calculations [Eq.\ (\ref{eproj}), solid 
line (online green)] for $N=17$ electrons at high magnetic field.
Right: The corresponding classical (rigid rotor) energy $E^{\text{rig}}_L$ 
for $N=17$ electrons (see text). The microscopic REC energies are referenced 
relative to the zero-point energy, $17 \hbar \Omega$. Energies 
were calculated for magic angular momenta 
$L=L_1+L_2+L_3$ with $L_1=0$, $L_2=21+6k_2$ and $k_2=30$, and $L_3=115+10k_3$.
The parameters are the same as in Fig.\ 1 and $B=100$ T.
Note the much larger energy scale for the classical case (right frame),
leading to a superfluidity index for the REC of $\alpha_s \sim
0.99$ (see text).
}
\end{figure}

In Fig.\ 3 (left frame), and for a sufficiently high magnetic field 
(e.g., $B=100$ T such that the Hilbert space of the system reduces to the
lowest Landau level), we compare the approximate analytic energies 
$E_{\text{app},L}^{\text{REC}}$ with the microscopic energies
$E_L^{\text{REC}}$ calculated from Eq.\ (\ref{eproj}) using the same 
parameters as in Fig. 1. The agreement is very good (typically less
than 0.5\%). More important is the strong discrepancy between these results and
the total energies $E^{\text{rig}}_L$ of the {\it classical\/} 
(rigid rotor) crystallite, plotted in the right frame of Fig.\ 3; the latter 
are given by $E^{\text{rig}}_L = \hbar^2 L^2 / (2 {\cal J}_{\text{rig}}) 
+ 0.5 \sum_{i=1}^N m^* \omega_0^2 |Z_i|^2
+ \sum_{i=1}^N \sum_{j > i}^N e^2/ (\kappa |Z_i-Z_j|)$, with the rigid moment 
of inertia being 
${\cal J}_{\text{rig}}=\sum_{i=1}^N m^* |Z_i|^2$ \cite{note3}.

The disagreement between the REC and classical energies is twofold: (i) 
The $L$ dependence is different, and (ii) The REC energies are three orders of 
magnitude smaller than the classical ones. That is, the energy cost for the
rotation of the REC is drastically smaller than for the classical rotation, 
thus exhibiting ``superfluid'' behavior. In analogy with Ref.\ \cite{legg}, we 
define a supefluidity index 
$\alpha_s = (E^{\text{rig}}_L - E_L^{\text{REC}})/E^{\text{rig}}_L$. For the
case displayed in Fig.\ 3, we find that this index varies 
(for $1116 \leq L \leq 3716$) from $\alpha_s = 0.978$ to $\alpha_s = 0.998$, 
indicating that the REC is highly superfluidic.

Formation of a supersolid is often expected in conjunction with the presence 
of (i) real defects and (ii) real vacancies \cite{anlf,ches}. Our supersolid 
wave function [Eq.\ \ref{wfprj}] belongs to a third possibility, namely to 
{\it virtual\/} defects and vacancies when the number of particles 
equals the number of lattice sites (the possibility of a supersolid with equal
number of particles and lattice sites is mentioned in Ref.\ \cite{legg2}). 
Indeed, the azimuthal shift of the electrons by 
$(\gamma_1, \gamma_2,...,\gamma_r)$ [see Eq.\ (\ref{wfprj})] may be viewed 
as generating {\it virtual\/} defects and vacancies with respect to the original
electron positions at $(\gamma_1=0, \gamma_2=0,...,\gamma_r=0)$. 

A recent publication \cite{jeon} has explored the quantal nature of the
2D electron crystallites in the lowest Landau level ($B \rightarrow \infty$)
using a modification of the second-quantized LLL form of the REC wave 
functions \cite{yl1}. In particular, the modification
consisted of a multiplication of the {\it parameter free\/} REC wave function by
variationally adjustable Jastrow-factor vortices. Without consideration of the 
rotational properties of the modified wave function, the inherently quantal 
nature of the crystallite was attributed exclusively to the Jastrow factor. 
However, as shown above, the original REC wave function [Eq.\ (\ref{wfprj})] 
already exhibits the characteristic NCRI behavior of a supersolid. 
Consequently, the additional {\it variational\/} freedom introduced by the 
Jastrow prefactor may well lead energetically to a slight numerical improvement,
but it does not underlie the physics related to the supersolid behavior.
Indeed, the essential step is the projection of the static electron crystallite 
onto a state with good total angular momentum [see Eqs. (\ref{uhfo}) and 
(\ref{wfprj})]. 

In summary, we have shown that rotating electron crystallites, formed in
2D quantum dots under high magnetic fields, exhibit nonclassical rotational
inertia which is the signature of a supersolid \cite{legg,legg2}.
We have utilized an analytic many-body wave function [Eq.\ (\ref{wfprj})]
which allowed us to carry out computations for a sufficiently large
number of electrons ($N=17$ electrons having a nontrivial three-ring
polygonal structure), leading to the derivation and validation of an analytic 
expression for the total energy of rotating electron crystallites of arbitrary
$N$. These crystallites may be regarded as the precursors of an extended 2D 
fermionic supersolid in the lowest Landau level.

This research is supported by the US D.O.E. and the NSF.


\begin{thebibliography}{99}
\bibitem{note1}
The REC is also referred to as rotating electron molecule
(REM) or rotating Wigner molecule (RWM).
\bibitem{note3}
For the classical equilibrium structure of $N=17$ point charges, the inner
radius is $0.9696R_0$, while the outer one is $1.8418R_0$ \cite{kon}.
\end{thebibliography}
\end{document}